\documentclass[reprint, secnumarabic, amssymb, amsmath, superscriptaddress, aps, prx]{revtex4-2}
\usepackage{graphicx}
\usepackage[colorlinks=true,urlcolor=blue]{hyperref}
\usepackage{xcolor}

\hypersetup{
colorlinks,
linkcolor={red!50!black},
citecolor={blue!50!black},
urlcolor={blue!80!black}
}
\usepackage[normalem]{ulem}
\usepackage{units}

\begin{document}

\title{Fundamental Tests of Quantum Geometric Bounds in Ionic and Covalent Insulators using Inelastic X-Ray Scattering}

\author{David Bałut}
\affiliation{Department of Physics, University of Illinois, Urbana, Illinois 61801}

\author{Barry Bradlyn}
\affiliation{Department of Physics, University of Illinois, Urbana, Illinois 61801}
\affiliation{Anthony J Leggett Institute for Condensed Matter Theory, University of Illinois, Urbana, Illinois 61801}

\author{Marcus D. Collins}
\affiliation{Astera Institute}

\author{Peter Abbamonte}
\affiliation{Department of Physics, University of Illinois, Urbana, Illinois 61801}
\affiliation{Materials Research Laboratory, University of Illinois, Urbana, Illinois 61801}
\email{abbamonte@mrl.illinois.edu}

\begin{abstract}

Quantum geometry underlies many fundamental properties of materials, but
it has remained largely inaccessible to direct experiment. Here we demonstrate that inelastic x-ray scattering (IXS) provides a direct, quantitative probe of quantum geometry and quantum information in solids. Studying two prototype insulators, covalently bonded diamond and ionically bonded LiF, we measure the density response and experimentally determine the quantum Fisher information, the associated Bures metric, and the electron localization length. These measurements enable a quantitative comparison of quantum geometry for two distinct bonding environments. We find that the dimensionless quantum weight, $aK(q)$, which quantifies the longitudinal localization of quantum information, is constrained by fundamental electrostatic bounds in both materials. Crucially, the quantum weight of diamond exceeds that of LiF, indicating that covalent bonds exhibit a higher degree of delocalization and higher density of quantum information than the ionic bonds. Our results establish a direct experimental relationship between quantum information, electron localization, and chemical bonding, and identify IXS as a powerful tool for measuring quantum geometry in materials.

\end{abstract}

\maketitle

\section{Introduction}
It has recently become clear that the geometry of quantum states plays a fundamental role in determining material properties. The first clear manifestation of quantum geometry arose in the quantum Hall effect, where the quantized Hall conductivity of a two-dimensional system can be expressed as an integral of the Berry curvature over the space of twisted boundary conditions (fluxes)~\cite{niu1985quantized}. The Berry curvature, however, represents only the imaginary part of a more general quantity called the quantum geometric tensor (QGT) whose real part, the quantum metric, is symmetric on the space of quantum states~\cite{provost1980riemannian} and quantifies the ground state polarization fluctuations at zero temperature~\cite{souza2000polarization}.

It was recently established that, while the Berry curvature determines the nondissipative Hall response, the quantum metric is related through a sum rule to the dissipative component of the longitudinal conductivity~\cite{souza2000polarization}. This same twisted boundary condition metric is also a measure of the amplitude of polarization fluctuations in insulators, and therefore provides a measure of electronic localization. In weakly interacting systems, the quantum metric associated with single-particle wavefunctions has been shown to play a central role in stabilizing exotic interacting topological phases, controlling the localization of Wannier functions, and, due to the connection to twisted boundary conditions, to place constraints on the phase stiffness of superconductors~\cite{torma2023essay,torma2022superconductivity,yu2025quantum,parameswaran2013fractional,marzari1997maximally}. More broadly, the unification of the Berry curvature and quantum metric within the QGT implies that the Berry curvature sets a lower bound on the metric, and hence a fundamental lower bound on localization in insulators~\cite{ozawa2021relations}.

The quantum geometric interpretation of the zero-temperature conductivity in insulators represents only a special case of a much broader connection between linear response functions and quantum geometry~\cite{verma2025quantum,ji2025density,guanExploringManyBodyQuantum2025,hetenyi2023fluctuations}. By leveraging the information-theoretic geometry of quantum states, this connection can be extended to finite temperatures, more general parameter spaces, and even to metallic systems. 

At nonzero temperature, a material in equilibrium is described by a thermal density matrix, $\rho_\beta$, with inverse temperature, $\beta$. To probe the system experimentally, one applies a time-dependent external perturbation, driving the system into a nonequilibrium state described by a time-dependent density matrix, $\rho(t)$. A natural question to ask is to what degree $\rho(t)$ differs from the equilibrium $\rho_\beta$. An information-theoretic measure of this difference is given by the Bures distance~\cite{Bengtsson_Zyczkowski_2006},
\begin{equation}
d_B(\rho_\beta,\rho(t))^2=2\left [ 1-\sqrt{F(\rho_\beta,\rho(t))} \right ]
\end{equation}
which is defined in terms of the fidelity
\begin{equation}
\sqrt{F(\rho_\beta,\rho(t))} = \mathrm{tr} \sqrt{\sqrt{\rho_\beta}\rho(t)\sqrt{\rho_\beta}}. 
\end{equation}
The fidelity is a quantitative measure of the similarity between the density matrices $\rho_\beta$ and $\rho(t)$. More explicitly, the fidelity corresponds to the maximum probability, over all possible measurements, of obtaining the state $\rho(t)$ from a measurement performed on $\rho_\beta$ through its coupling to some external environment \cite{uhlmann1976transition}. The Bures distance therefore serves as a quantitative measure of the distinguishability, or dissimilarity, between equilibrium and nonequilibrium quantum states.

For weak perturbations, we expect that $\rho(t)\approx \rho_\beta + \rho_1(t)$, where $\rho_1(t) \ll \rho_\beta$ is first order in a perturbation expansion. Expanding the Bures distance to leading order in $\rho_1(t)$ gives  
\begin{equation}
d_B(\rho_\beta,\rho(t)) \approx \frac{1}{4}F_Q(t),
\end{equation}
where $F_Q(t)$ is the quantum Fisher information (QFI) associated with the perturbation $\rho_1(t)$. 
The quantum Fisher information quantifies how well we can distinguish $\rho_\beta$ and $\rho_\beta+\rho_1(t)$ given $N$ arbitrary measurements. That is, the quantum Cram\'er-Rao bound implies that $1/[NF_Q(t)]$ sets the minimum possible uncertainty 
of any unbiased estimator of the difference between the two density matrices~\cite{braunstein1994statistical}.

The quantum Fisher information provides a natural way to generalize the zero temperature twist angle metric for the conductivity to nonzero temperature and to general perturbations. The second derivative of $F_Q(t)$ with respect to the external field defines the Bures metric associated with the perturbation. For a time-independent transverse electric field, the external field corresponds to a twisted boundary condition, recovering the usual twist-angle metric at zero temperature. Considering different dynamical perturbing fields allows one to define the Bures metric on a variety of parameter spaces.

The key unifying principle is that the Bures metric associated with a perturbation coupling to operators $B_\mu$ can be expressed directly in terms of the system’s linear response to that perturbation~\cite{haukeMeasuringMultipartiteEntanglement2016}.
In terms of the frequency components of the applied field, the Bures metric takes the form
\begin{equation}\label{eq:information-dissipation}
g_{\mu\nu}(\omega) = -\frac{1}{2\pi}\tanh\frac{\beta\omega}{2}\chi''_{\mu\nu}(\omega),
\end{equation}
where $\chi''_{\mu\nu}(\omega)$ is the antihermitian part of the response function for the operator $B_\mu$ ~\cite{kadanoff1963hydrodynamic}. 
Through the fluctuation–dissipation theorem, $\chi''_{\mu\nu}(\omega)$
is proportional to the equilibrium fluctuations of the operators $B_\mu$. Eq.~\eqref{eq:information-dissipation} shows that these same fluctuations are \emph{also} proportional to the Fisher information encoded in the perturbation $B_\mu$. Viewed another way, Eq.~\eqref{eq:information-dissipation} tells us that the change in information induced by a perturbation (quantified by the Bures metric $g_{\mu\nu}$) is proportional to the rate at which the perturbation drives transitions, as described by Fermi’s golden rule. The value of the Bures metric is therefore given by 
$\chi''_{\mu\nu}(\omega)$ multiplied by the relative change in thermal occupation probability between the initial and final states, captured by the factor $\tanh{\frac{\beta \omega}{2}}$.

{If $B_\mu$\ is chosen to be the scalar density operator, $\hat{\rho}(r,t)$, the quantum Fisher information becomes an experimentally measurable probe of quantum geometry. The response function entering the Bures metric in Eq.~\eqref{eq:information-dissipation} is then the density–density response function, $\chi''(\mathbf{q},\omega)$, which can be directly measured with inelastic x-ray scattering (IXS) \cite{schuelkeElectronDynamicsInelastic2007}. In quantum geometry, the primary quantity of interest is the Bures metric due to instantaneous perturbations \cite{haukeMeasuringMultipartiteEntanglement2016,balutQuantumEntanglementQuantum2025}, and is obtained by integrating Eq.~\eqref{eq:information-dissipation} over frequency,}
\begin{equation}\label{eq:densitymetric}
g_\mathbf{q} = -\frac{\hbar}{\pi}\int_0^\infty d\omega \tanh\frac{\beta\omega}{2}\chi''(\mathbf{q},\omega). 
\end{equation}
This version of the Bures metric $g_\mathbf{q}$ is proportional to the quantum Fisher information density $f_Q$ of Refs.~\cite{haukeMeasuringMultipartiteEntanglement2016,balutQuantumEntanglementQuantum2025,balut2025quantum,fang2025amplified,wang2025local,mazza2024quantum}:
\begin{equation}
4g_\mathbf{q} = f_Q.\label{eq: QFI sum rule}
\end{equation}

This quantity provides a direct measure of charge localization in insulators, and constitutes a finite-temperature generalization of the longitudinal part of the quantum metric, applicable to both insulators and metals. At small $\mathbf{q}$, because $\rho = - \nabla \cdot \mathbf{P} \sim \mathbf{q}\cdot \mathbf{P}$, the density response, $\chi''(\mathbf{q},\omega)$, represents the spectrum of longitudinal polarization fluctuations. Accordingly, $g_\mathbf{q}$ quantifies the degree to which valence electrons are displaced relative to their associated ionic cores, through the sensitivity of $\chi''(\mathbf{q},\omega)$ to the geometry of the many-body density matrix. In the zero-temperature limit, the Bures metric in Eq.~\eqref{eq:densitymetric} reduces to the longitudinal component of the quantum metric. Taken together, the longitudinal and transverse parts of the quantum metric characterize charge delocalization in insulators, the former capturing screening due to long-ranged Coulomb interactions~\cite{resta2006polarization} and the latter characterizing the twist-angle dependence of the ground state.

Longitudinal polarization fluctuations have also recently been discussed in terms of the so-called {\it quantum weight}~\cite{onishiQuantumWeightFundamental2025}, defined for insulators as the small-$\mathbf{q}$, zero-temperature limit of the static structure factor,
\begin{equation}\label{Kdef}
    K^{\mu\nu} = \left.\pi\frac{\partial S(\mathbf{q})}{\partial q_\mu\partial q_\nu}\right|_{\mathbf{q=0},T=0}.
\end{equation}
For an insulator at zero temperature, the static structure factor is given by,
\begin{equation}\label{eq:zeroTssf}
    S(\mathbf{q})\rightarrow -\frac{\hbar}{\pi}\int_0^\infty d\omega \chi''(\mathbf{q},\omega).
\end{equation}
Using Eq.~\eqref{eq:densitymetric}, we can therefore relate the quantum weight to the Bures metric via
\begin{equation}\label{quantumweightdef}
K^{\mu\nu} = \left.\pi\frac{\partial^2 g_\mathbf{q}}{\partial q_\mu\partial q_\nu}\right|_{\mathbf{q}=T\rightarrow 0}.
\end{equation}
This shows that the quantum weight measures the information encoded in longitudinal charge fluctuations in insulators at low temperatures. As such, it provides direct insight into the nature of chemical bonding in insulators. A convenient dimensionless measure of charge localization is given by $aK^{\mu\nu}$, where $a$ is the lattice constant. In ionically bonded materials, we expect electrons to be strongly localized near the negative ions with relatively
small position fluctuations, leading to $aK^{\mu\nu}<1$. By contrast, electrons in covalently bound (i.e. obstructed atomic~\cite{bradlyn2017topological,komissarov2024quantum}) insulators are more delocalized and fluctuate between different atomic sites, leading us to expect $aK^{\mu\nu}>1$. This reflects the fact that, in covalent solids, quantum information is shared across multiple unit cells.

To test these ideas experimentally, we use IXS to measure the quantum weight for two prototypical insulators: the ionic rock salt LiF and covalently bonded diamond. We use IXS data to compute the Bures metric and the $\mathbf{q}$-dependent longitudinal quantum weight 
\begin{equation}\label{eq:Kq from g}
    K(\mathbf{q})\equiv \frac{2\pi}{q^2}g_\mathbf{q}
\end{equation}
for both LiF and diamond, with $q=|\mathbf{q}|$. In the small $q$ limit, Eq.~\eqref{quantumweightdef} shows that $K(\mathbf{q})$ reduces to the longitudinal component of the quantum weight,
\begin{equation}\label{eq: Kq small q limit}
    K(\mathbf{q}\rightarrow 0) = \frac{q_\mu q_\nu}{q^2}K^{\mu\nu}.
\end{equation}

We take the approach of using the $f$-sum rule to normalize the experimental data, enabling a quantitative comparison of $K(\mathbf{q})$ between the two materials. For LiF, our analysis refines and extends the proof-of-principle calculation first presented in Ref.~\cite{balutQuantumEntanglementQuantum2025}. 
Generalizing the results of Ref.~\cite{onishiQuantumWeightFundamental2025}, we derive wavevector-dependent upper and lower bounds on the Bures metric, demonstrating that the quantum geometry of the density response is constrained by the electrostatic properties of insulators.
In particular, we will show in Section III that for any insulator
\begin{equation}
    (1 - \epsilon^{-1}(\mathbf{q}))E_g \leq 4e^2 K(\mathbf{q}) \leq \sqrt{1 - \epsilon^{-1}(\mathbf{q})} \hbar \omega_p,
    \label{eq: K Bound}
\end{equation}
where $E_g$ is the smallest gap to electronic excitations, $\epsilon(\mathbf{q}) = \mathrm{Re}~\epsilon(\mathbf{q}, \hbar \omega < E_g$) is the electronic contribution to the static longitudinal dielectric function, and $\omega_p \equiv \sqrt{4\pi n e^2/m}$ is the plasma frequency defined in terms of the bare electron mass $m$ and total electron density $n$. 
{In the limit $\mathbf{q}\rightarrow 0$, Eq.~\eqref{eq: K Bound} reduces to the bounds on the uniform quantum weight Eq.~\eqref{Kdef} derived in Ref.~\cite{onishiQuantumWeightFundamental2025}.}
Crucially, we show how by applying both sum rules and the Kramers-Kronig relations, all quantities entering these bounds can be determined self-consistently from experimental data alone.
We show how truncated, partial sum rules can be used to estimate $\omega_p$
from the finite frequency window accessible to IXS experiments. We also show that numerical Kramers-Kronig relations allow us to determine the static $\epsilon(\mathbf{q})$ from the measured spectra. 
Taken together, these results show how the Bures metric can be used to probe the nature of chemical bonding using x-ray scattering.

Additionally, we explore the connection between the Bures metric and quantum entanglement. For free fermion systems, we show that the Bures metric is lower bounded by a function of the bipartite entanglement eigenvalues. By analyzing simple tight-binding models, we demonstrate that the Bures metric directly probes the entanglement inherent to covalent bonds. 
Finally, we compare the experimentally extracted Bures metric for diamond with the corresponding quantity computed within a minimal tight-binding model, finding good qualitative agreement.

\section{Experiment}\label{sec:experiment}

IXS experiments were done at room temperature at Sector 9-ID of the Advanced Photon Source using a liquid-nitrogen-cooled Si(111) primary monochromator and Si(333) secondary channel-cut. Diamond data were collected at 8981.8 eV using a Ge(733) backscattering analyzer (0.3 eV resolution) with $q \parallel (1,\bar{1},0)$, while LiF data were collected at 6482.5 eV using a Ge(531) analyzer (140 meV resolution) with $q \parallel (1,0,0)$. All measurements had momentum resolution $\Delta q \sim 0.15$ \AA$^{-1}$, and energy-loss scans were obtained by scanning the incident energy. 

A key advantage of IXS is that it directly measures the density response function, enabling quantitative determination of the quantum Fisher information, Bures metric, and quantum weight $K^{\mu\nu}$ via Eqs.~\eqref{eq: QFI sum rule}, \eqref{eq:densitymetric}, and \eqref{eq: Kq small q limit}.
The x-ray differential scattering cross section is given by \cite{schuelkeElectronDynamicsInelastic2007}
\begin{equation}
    \frac{\partial^2 \sigma}{\partial \Omega \partial \omega} = r_0^2 \frac{\omega_f}{\omega_i} \left | \epsilon_f^* \cdot \epsilon_i \right |^2 S(\mathbf{q},\omega),
    \label{eq: IXS Cross Section}
\end{equation}
where $r_0$ is the classical electron radius, $\epsilon_i$ and $\epsilon_f$ are the incident and scattered polarizations, {and the ratio of initial to final photon frequencies is} $\omega_i / \omega_f \approx 1$. 
The excitation spectrum is encoded in the dynamic structure factor $S(\mathbf{q},\omega)$, the density–density correlation function, which is related to the dissipative part of the density response $\chi(\mathbf{q},\omega)$ through the fluctuation dissipation theorem:
\begin{equation}
    S(\mathbf{q},\omega) = \frac{-1}{\pi}\frac{1}{1 - e^{-\hbar \omega/k_bT}} \chi''(\mathbf{q},\omega).
    \label{eq: fluctuation dissipation theorem}
\end{equation}
In this study, we focus on electronic excitations above the band gaps of diamond (5.4 eV) and LiF (14 eV). At these high energy scales, the Bose factor in Eq.~\eqref{eq: fluctuation dissipation theorem} is effectively unity at room temperature. Consequently, to extract the susceptibility from the measured spectra, we subtract the elastic scattering contribution and assume that the resulting signal vanishes at energies well below the band gap. This treatment of elastic scattering differs from our previous work~\cite{balutQuantumEntanglementQuantum2025}, in which the spectra were antisymmetrized. The approach adopted here fully removes elastic interference and yields a more accurate determination of the quantum Fisher information, quantum weight, and related quantities in this case.

\begin{figure}
    \centering
    \includegraphics[width=\linewidth]{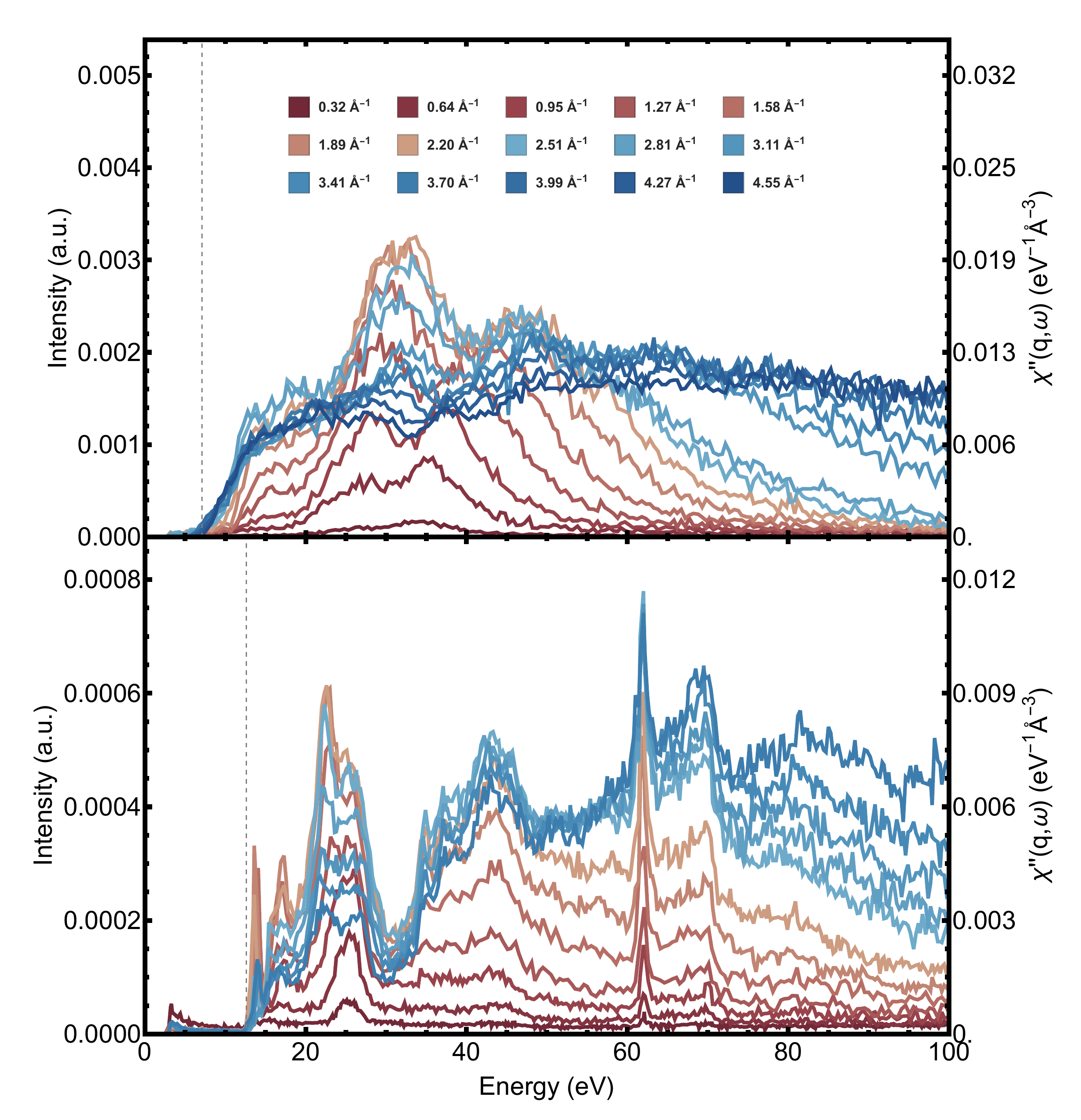}
    \caption{IXS spectra of diamond (top) and LiF (bottom) for various momentum transfer. The quasi-elastic peak, corresponding to data below $3 \ eV$, have been set to zero. The vertical lines correspond to the $q \rightarrow 0$ value of the band gap as reported in~\cite{onishiQuantumWeightFundamental2025}.}
    \label{fig: raw data plot}
\end{figure}

IXS spectra for both LiF and diamond, measured up to an energy loss of 100 eV for a selection of momenta, $q$, are shown in Fig.~\ref{fig: raw data plot}. In LiF, the spectra comprise (listed in order of increasing energy) a sharp exciton line, a continuum of interband transitions, shallow core-level excitations, the Li K edge, and, at large momentum transfers, and the emergence of a broad Compton profile, as reported previously~\cite{Caliebe2000,AbbamonteLiF2008,Lee-LiF-2013,balut2025quantum}.

At smaller momentum transfers, the diamond spectra exhibit three distinct features whose microscopic origin is unclear. It is tempting to interpret them as collective excitations involving one, two, and three electron-hole pairs promoted across the band gap, constituting a plasmon-like collective response in this wide-gap insulator. At larger momentum transfers, these features broaden and, as in the case of LiF, eventually evolve into a Compton profile.

Even with a detailed understanding of the microscopic origin of all spectral features, inspection of Fig.~\ref{fig: raw data plot} alone does not provide any concrete conclusions about the Fisher information or information geometry of these materials. Such conclusions can only be drawn by applying Eqs.~\eqref{eq: QFI sum rule},~\eqref{eq:densitymetric}, and \eqref{eq: Kq small q limit} to the IXS data in order to extract the Bures metric and the quantum weight.

Doing so requires addressing two practical limitations. First, the spectra shown in Fig.~\ref{fig: raw data plot} represent the photon count rate, which is proportional to the IXS cross section, Eq.~\eqref{eq: IXS Cross Section}. These spectra must be properly scaled to obtain the susceptibility, $\chi''(\mathbf{q},\omega)$, in physical units of $E^{-1}L^{-3}$. Second, the IXS spectra were measured over a large but finite frequency range, which prevents Eq.~\eqref{eq:densitymetric} from being integrated to $\omega \rightarrow \infty$. Both limitations can be overcome through judicious application of the $f$-sum rule, as we now show.

\subsection{Extrapolation and Scaling using the Sum Rule}\label{sub:extrapolation_and_scaling_using_the_sum_rule}

The $f$-sum rule relates the first frequency moment of the response function $\chi''(\mathbf{q},\omega)$ to the electron density $n$ via \cite{Martin1968,schuelkeElectronDynamicsInelastic2007,balutQuantumEntanglementQuantum2025}
\begin{equation}
    \int_0^{\infty}d \omega \omega \chi''(\mathbf{q},\omega) = -\frac{\pi n q^2}{2m},
    \label{eq: f-sum rule}
\end{equation}
where $m$ is the free electron mass. 
If $\chi''(\mathbf{q},\omega)$ were known at all frequencies and the integral could be evaluated exactly, then $n$ would equal the total electron density of the material, thereby providing the exact scale for $\chi''(\mathbf{q},\omega)$ in absolute units.

In practice, $\chi''(\mathbf{q},\omega)$ can be measured only up to a finite maximum frequency $\Omega$, which in our experiment is 100~eV.
In this case, it has been shown that when $\Omega$ lies within an energy gap separating well-resolved subshells of excitations, Eq.~\eqref{eq: f-sum rule} can be reformulated as a {\it partial} sum rule, with $n$ replaced by $n^*/\epsilon_\infty$, with $n^*$ being an effective density that counts only the electrons in a subset of shallow shells, and $\epsilon_\infty$ describing background screening due to excitations above $\Omega$
\cite{Ehrenreich1963,Taft1965,Shiles1980}.

For diamond, our cutoff of 100~eV includes all valence electron excitations but lies well below the carbon K edge at 280~eV. The corresponding effective density is therefore four electrons per carbon atom, or eight per unit cell. Because K-edge excitations occur at much higher energies, their contribution to screening is negligible, implying $\epsilon_\infty \simeq 1$ in this case.

In LiF, the fluorine atom has seven active valence electrons below the cutoff energy, while lithium contributes two valence electrons and one shallow core electron due to excitation of the Li K edge near 65~eV. This yields an effective  density of ten electrons per unit cell. {As in diamond, the cutoff $\Omega$ is well below the fluorine K-edge at 697eV, so there is again negligible screening from the core electrons implying $\epsilon_\infty\simeq 1$.} Using the experimentally determined (primitive) lattice constants, $a_{\mathrm{LiF}} = 2.84$~\AA ~\cite{PrecisionMeasurementLattice} and $a_{\mathrm{diamond}} = 2.52$~\AA ~\cite{straumanis_precision_1951}, gives the effective electron density for each material in \AA$^{-3}$. 

An additional complication is that, for momenta larger than about 2~\AA$^{-1}$, the experimental intensity in Fig. 1 does not vanish above {the cutoff $\Omega=100$} eV due to the emergence of a Compton profile. 
Hence, at these momenta, the IXS data do not capture the full spectral weight of even the effective electrons. To account for the missing weight, the spectra must be extrapolated to $\omega \rightarrow \infty$. The $f$-sum rule, Eq.~\eqref{eq: f-sum rule}, constrains the high-frequency asymptotic behavior of $\chi''(\mathbf{q},\omega)$.
Assuming a polynomial decay at large $\omega$, which is reasonable for an insulator, the response should scale as
\begin{equation}
\chi''(\mathbf{q},\omega\rightarrow\infty)\sim \mathcal{O}(\omega^{-3}).
\end{equation}
We therefore extrapolate the IXS spectra in Fig.~\ref{fig: raw data plot} according to
\begin{equation}
{I(\mathbf{q},\hbar\omega>\Omega) = \frac{c(\mathbf{q})\Omega^3}{(\hbar\omega)^3},}
\end{equation}
where $c(\mathbf{q})$ is the average spectral intensity between 94 and 100~eV at fixed momentum. With this extrapolation, we can evaluate the integral in Eq.~\eqref{eq: f-sum rule} explicitly.

\begin{figure}
    \centering
    \includegraphics[width=\linewidth]{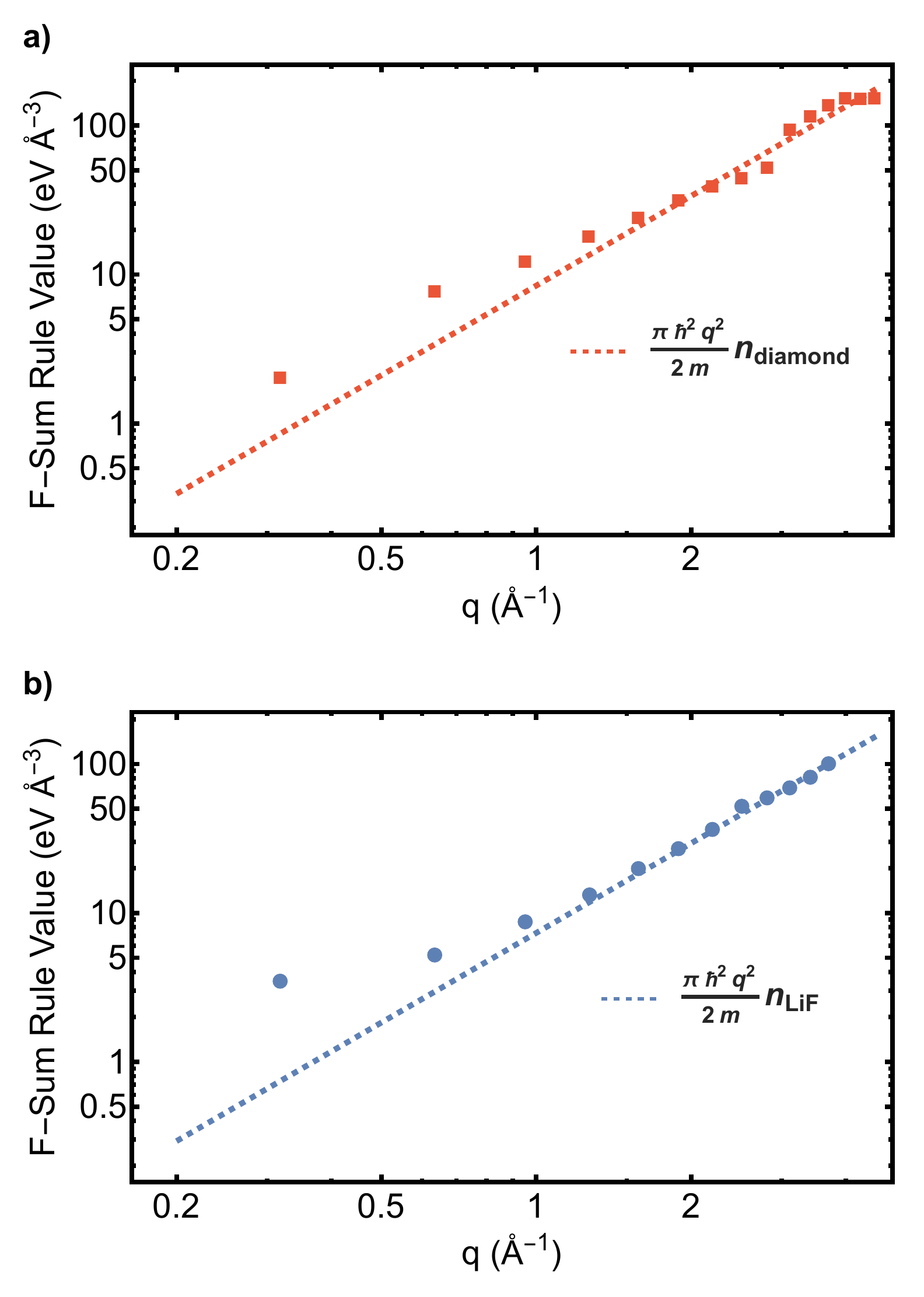}
    \caption{Value of the $f$-sum rule integral [Eq.~\eqref{eq: f-sum rule}] for the scaled IXS data for (a) diamond and (b) LiF, plotted against the momentum transfer. The dashed line represents the right hand side of Eq.~\eqref{eq: f-sum rule} evaluated using the effective electron density. 
    Both data sets were scaled with a momentum-independent constant obtained from a $q^2$ fit to the high-momentum spectra.}
    \label{fig: f sum rule}
\end{figure}

The resulting values of the $f$-sum rule integral, Eq.~\eqref{eq: f-sum rule}, which summarize our chosen scaling for both the diamond and LiF datasets, are shown in Fig.~\ref{fig: f sum rule}. Overall, the spectra display the expected $q^2$ dependence, except at the lowest momenta where excess spectral weight is observed. This additional weight arises from the finite momentum resolution of the measurements, as discussed previously~\cite{balutQuantumEntanglementQuantum2025}.
Accordingly, when scaling the data we excluded the four lowest momentum points in diamond and the three lowest in LiF, placing greater weight on the higher-momentum regime.

\section{Quantum Fisher information for diamond and LiF} 

\begin{figure}
    \centering
    \includegraphics[width=\linewidth]{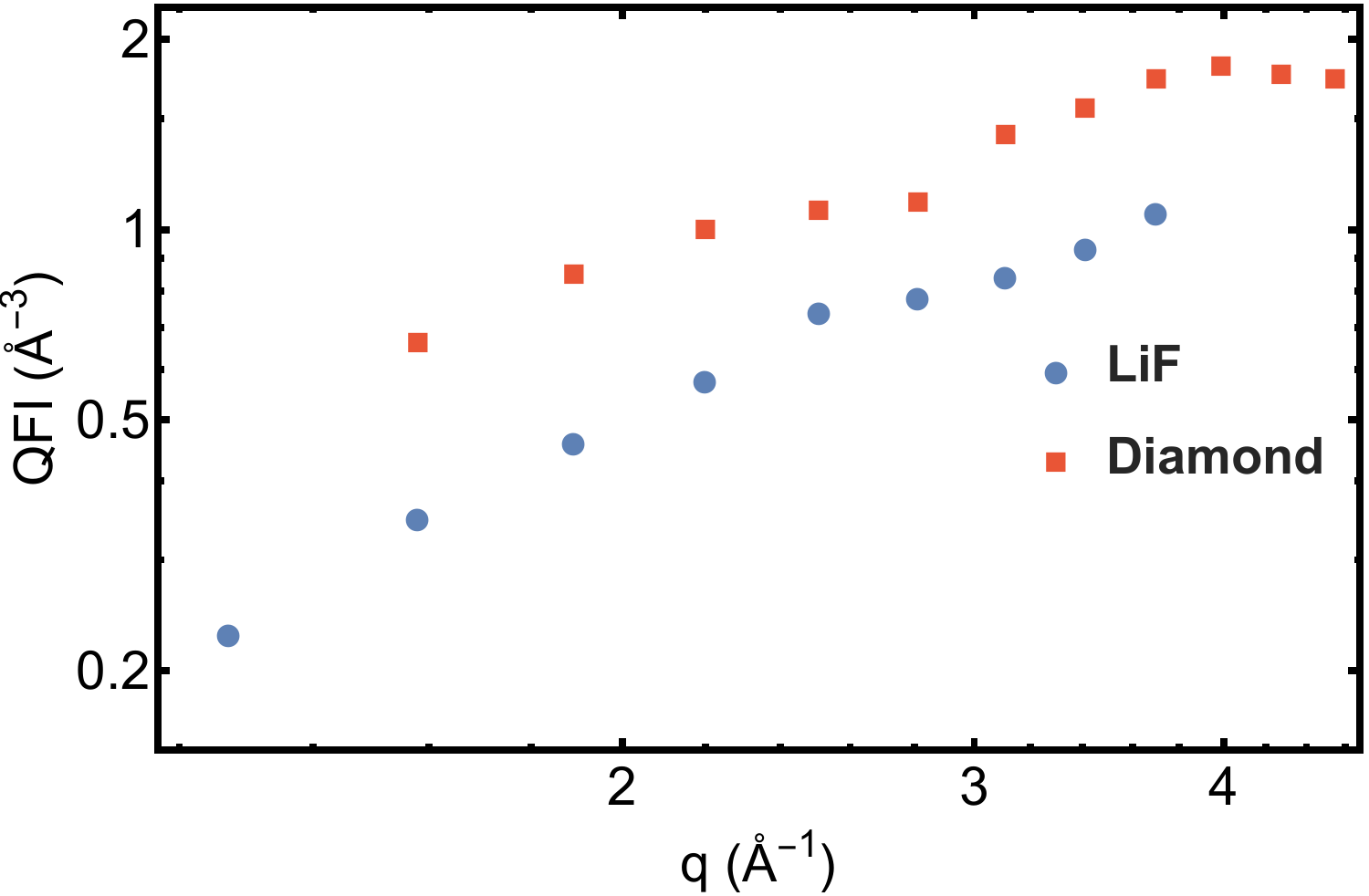}
    \caption{The QFI as a function of $q$ from IXS data on log-log scale. }
    \label{fig: QFI result}
\end{figure}

We now determine the Bures metric and quantum weight for LiF and diamond by computing the quantum Fisher information using Eqs.~\eqref{eq:densitymetric} and \eqref{eq: QFI sum rule}. The frequency integral in Eq.~\eqref{eq:densitymetric} is evaluated using the normalized and extrapolated response functions from Sec.~\ref{sec:experiment}. The results are shown on a log–log scale in Fig.~\ref{fig: QFI result}.

It is apparent from Fig.~\ref{fig: QFI result} that the QFI in diamond is larger than that in LiF for all $q$, indicating that density fluctuations in diamond have a higher information content, as expected for a covalently bonded solid. Further, as $q \to 0$, the experimentally measured QFI in LiF decreases approximately quadratically with $q$, as indicated by a slope near 2 on the log–log plot. This behavior demonstrates that our measurements access the small-$q$ regime, enabling an accurate determination of the quantum weight for LiF via Eq.~\eqref{eq:Kq from g} and a test of the bounds of Ref.~\cite{onishiQuantumWeightFundamental2025}, as shown in detail in  Sec.~\ref{sec:K-analysis}.

For the case of diamond, the QFI decreases more slowly in the small-$q$ regime, indicating that the experiment does not reach the true small-$q$ limit of density fluctuations. Consequently, when interpreting these results in terms of the quantum weight, any extrapolation to $q \to 0$ in diamond should be regarded as providing a lower bound on the zero-$q$ quantum weight.

\section{Constraints on the quantum weight}

To enable an unbiased comparison of the quantum Fisher information between materials with different electron densities, we introduce a dimensionless, $\mathbf{q}$-dependent generalization of the quantum weight of Ref.~\cite{onishiQuantumWeightFundamental2025},
\begin{equation}
aK(\mathbf{q}) = -\frac{2\pi a}{q^2}g_\mathbf{q} = -\frac{\pi a}{2q^2}f_Q.
\end{equation}
It was shown in Ref.~\cite{onishiQuantumWeightFundamental2025} that the zero-wavevector limit of $aK(\mathbf{q})$ in an insulator is bounded by its electrostatic properties, reflecting limits on longitudinal polarization fluctuations or, equivalently, on its degree of electronic delocalization. 
Here, we generalize these bounds to finite $\mathbf{q}$, placing fundamental constraints on both the quantum weight and the Bures metric.

\subsection{Derivation of wavevector-dependent bounds}

Following the approach of Ref.~\cite{onishiQuantumWeightFundamental2025}, we combine Eqs.~\eqref{eq:densitymetric} and \eqref{eq:Kq from g} to write the quantum weight as 
\begin{equation}\label{eq:Kqdef}
    aK(\mathbf{q})=-\frac{2a\hbar}{q^2}\int_0^\infty d\omega \chi''(\mathbf{q},\omega).
\end{equation}
To obtain a lower bound, we note that in an insulator the imaginary part of the susceptibility $\chi''(\mathbf{q},\omega)$ vanishes for $\hbar\omega < E_g$, where $E_g$ is the minimum gap to density excitations, which may correspond to an indirect transition at nonzero $\mathbf{q}$. As a result, for all $\omega$,
\begin{equation}
    -\chi''(\mathbf{q},\omega) \geq -\frac{E_g}{\hbar\omega}\chi''(\mathbf{q},\omega).\label{eq:K lower bound deriv}
\end{equation}
Integrating Eq.~\eqref{eq:K lower bound deriv} over frequency allows the right-hand side to be evaluated using the Kramers–Kronig relation,
\begin{equation}
   \chi'(\mathbf{q},\omega) = \frac{2}{\pi} \, \mathcal{P} \int_{0}^{\infty} \frac{\omega' \chi''(\mathbf{q},\omega')}{\omega'^2 - \omega^2} \, d\omega',
   \label{eq: kramers-kronig}
\end{equation}
which yields
\begin{align}
\int_0^\infty d\omega \frac{E_g}{\hbar\omega}\chi''(\mathbf{q},\omega)& = \frac{\pi E_g}{2\hbar}\chi'(\mathbf{q},0) \nonumber \\ 
&= -\frac{ E_g q^2}{8e^2\hbar}(1-\epsilon^{-1}(\mathbf{q}))\label{eq:kk-for-bound}
\end{align}
where we introduced the low-frequency electronic contribution to the longitudinal dielectric function $\epsilon^{-1}(\mathbf{q})$ defined as
\begin{equation}\label{eq:epsilon}
    \frac{1}{\epsilon(\mathbf{q},\omega)} = 1 + V(q)\chi(\mathbf{q},\omega),
\end{equation}
where $V(q) =4\pi e^2 / q^2$ is the Coulomb propagator. Combining Eqs.~\eqref{eq:Kqdef}, \eqref{eq:K lower bound deriv} and \eqref{eq:kk-for-bound}, we obtain the lower bound
\begin{equation}
    4e^2aK(q)\geq  aE_g(1-\epsilon^{-1}(\mathbf{q})).\label{eq:lowerbound}
\end{equation}

To obtain an upper bound for $aK(q)$ we make use of the Cauchy-Schwarz inequality. Specifically, we write
\begin{equation}
    -\chi''(\mathbf{q},\omega) = \sqrt{-\omega\chi''(\mathbf{q},\omega)}\sqrt{-\frac{\chi''(\mathbf{q},\omega)}{\omega}}.
\end{equation}
Using the Cauchy-Schwarz inequality combined with the definite sign of $\chi''(\mathbf{q},\omega)$, we find
\begin{align}
aK(q)&\leq \frac{2a\hbar}{q^2}\sqrt{\left(\int_0^\infty d\omega\;\omega\chi''(\mathbf{q},\omega)\right)\left(\int_0^\infty d\omega \frac{\chi''(\mathbf{q},\omega)}{\omega}\right)} \nonumber \\
&\leq \frac{a\hbar\omega_p}{4e^2}\sqrt{1-\epsilon^{-1}(\mathbf{q})},\label{eq:upperbound}
\end{align}
where the first integral has been expressed in terms of the plasma frequency $\omega_p$ using the $f$-sum rule [Eq.~\eqref{eq: f-sum rule}], and the second was expressed in terms of the static dielectric function by using the Kramers–Kronig relation [Eq.~\eqref{eq:kk-for-bound}].

Together, Eqs.~\eqref{eq:lowerbound} and \eqref{eq:upperbound} constitute a $\mathbf{q}$-dependent generalization of the bound derived in Ref.~\cite{onishiQuantumWeightFundamental2025}. The derivation of these expressions illustrates how these bounds can be applied directly to experimental data. As discussed in Sec.~\ref{sub:extrapolation_and_scaling_using_the_sum_rule}, we construct $\chi''(\mathbf{q},\omega)$ by extrapolating the measured IXS spectra to $\omega \rightarrow \infty$ using the $\omega^{-3}$ asymptotic form required by the $f$-sum rule. The overall normalization of $\chi''(\mathbf{q},\omega)$ is then fixed by enforcing the $f$-sum rule with an effective electron density $n^*$ determined by the experimental energy cutoff. With this empirically constructed response function, we can follow the steps used to obtain Eqs.~\eqref{eq:lowerbound} and \eqref{eq:upperbound}. Provided the value chosen for $E_g$ is consistent with the experimental data, the plasma frequency $\omega_p$ is taken to correspond to the effective density $n^*$, and the dielectric function $\epsilon^{-1}(\mathbf{q})$ is obtained directly via Kramers–Kronig relations, the resulting bounds on $K(\mathbf{q})$ depend only on quantities extracted from the experiment itself.

In Fig.~\ref{fig: raw data plot}, the vertical lines indicate the reported energy gaps $E_g$ for diamond and LiF from Ref.~\cite{onishiQuantumWeightFundamental2025}. All IXS spectra are approximately zero below this energy, indicating that our data are consistent with these values. In Sec.~\ref{sub:extrapolation_and_scaling_using_the_sum_rule}, we showed how $n^*$ (and hence $\omega_p$) can be determined from the extrapolated $f$-sum rule (Fig.~\ref{fig: f sum rule}).

\subsection{Static electronic dielectric constant from IXS} 

\begin{figure}
    \centering
    \includegraphics[width=\linewidth]{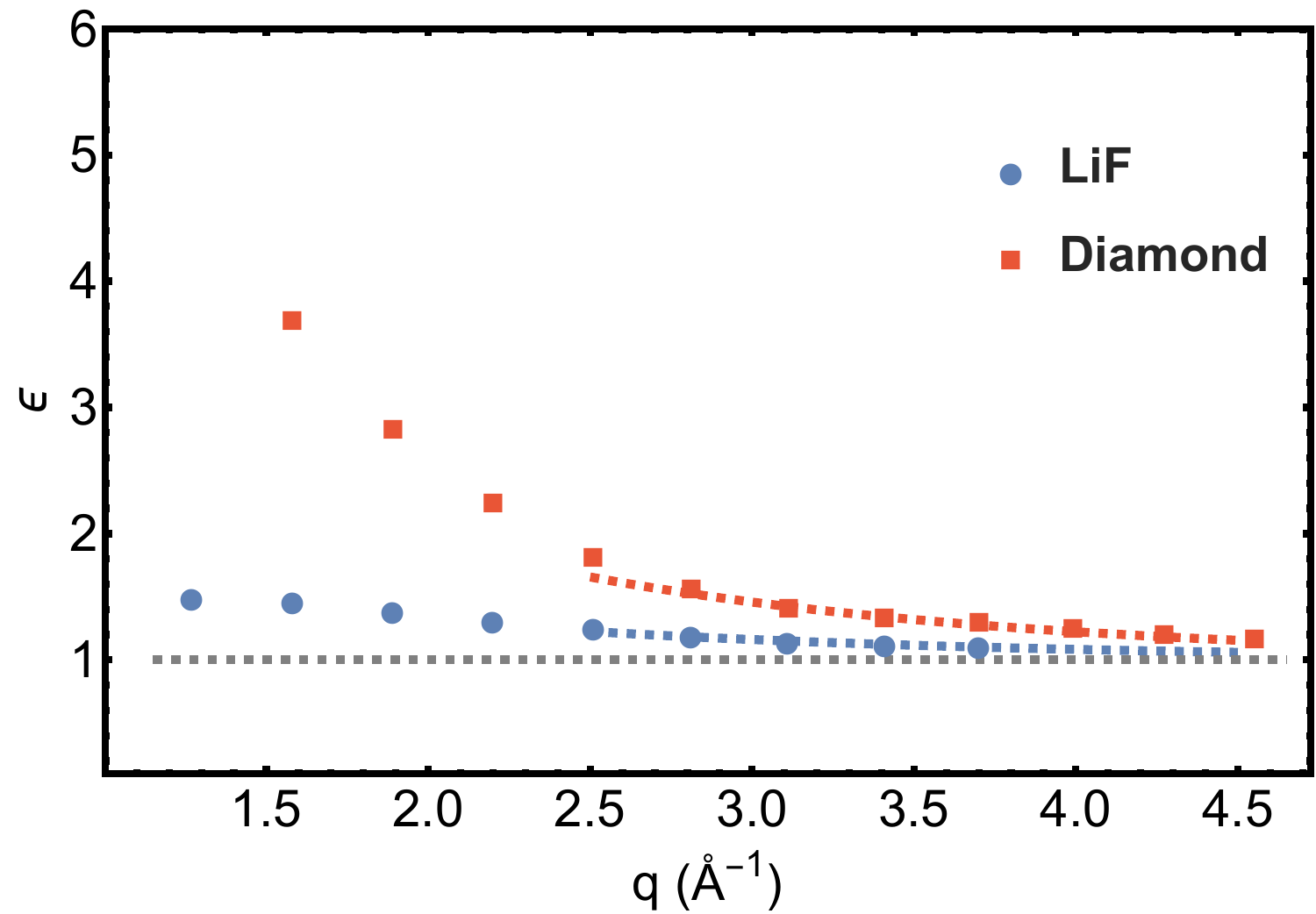}

    \caption{Result of using Kramers-Kronig on scaled IXS spectra and extracting $\epsilon(\mathbf{q)}$ below the gap ($4.5 \ eV$). The dashed lines correspond to the fit $\epsilon = 1 + e^{- \alpha q}$ for $q > 2.6$.}
    \label{fig: epsilon}
\end{figure}

The bound captured by Eq.~\eqref{eq: K Bound} is only useful if we have some way to determine the wave-vector dependent dielectric constant, $\epsilon(\mathbf{q})$. Here, we show that this quantity can be determined directly from IXS data. 
Specifically, we evaluate $\chi'(\mathbf{q},\omega)$ at a small but nonzero frequency within the gap, which isolates the electronic response by excluding contributions from phonons and extrinsic sample effects. Combining this result with Eq.~\eqref{eq:epsilon} yields the static electronic dielectric constant $\epsilon^{-1}(\mathbf{q})$. The resulting values are shown in Fig.~\ref{fig: epsilon}. Notably, our empirically determined electronic dielectric functions are consistent with the tabulated $\mathbf{q}\rightarrow 0$ limits $\epsilon_\mathrm{diamond}=5.7$ and $\epsilon_\mathrm{LiF}=1.96$~\cite{onishi2024fundamental}.

An informative consistency check on our computed values of $\epsilon(\mathbf{q})$ is provided by examining the large-$\mathbf{q}$ behavior. For $qa \gg 1$, the dielectric function is expected to decay exponentially to unity, with a decay rate set by the atomic length scale of the material,
\begin{equation}\label{eq:epsilon_form}
\epsilon(q) = 1 + e^{-\alpha q}.
\end{equation}
This reflects the fact that the very short-wavelength electric fields in a material are determined by the polarization of individual atoms, and are insensitive to long-wavelength collective effects.

To test this expectation, we fit the experimentally determined static dielectric function to Eq.~\eqref{eq:epsilon_form} over the range $q \ge 2.5$~\AA$^{-1}$, as shown in Fig.~\ref{fig: epsilon}. For diamond, this procedure yields a decay length $\alpha_{\mathrm{diamond}} = 0.72$~\AA, in excellent agreement with the known $sp^3$ covalent radius of carbon, $r_C \approx 0.76$~\AA \cite{cordero2008covalent}. Intriguingly, the decay length extracted for LiF, $\alpha_{\mathrm{LiF}} = 0.65$~\AA, is comparable to the covalent radius of fluorine, $r_F \approx 0.60$~\AA~\cite{robinson1997reinterpretation}. This agreement indicates that the short-wavelength electronic response of LiF is dominated by electrons localized on the fluorine atoms, consistent with the compound's ionic character.

\section{Dimensionless Quantum Weight and the Nature of Bonding}\label{sec:K-analysis}

\onecolumngrid\
\begin{center}\
    \begin{figure}
        \centering
        \includegraphics[width=\linewidth]{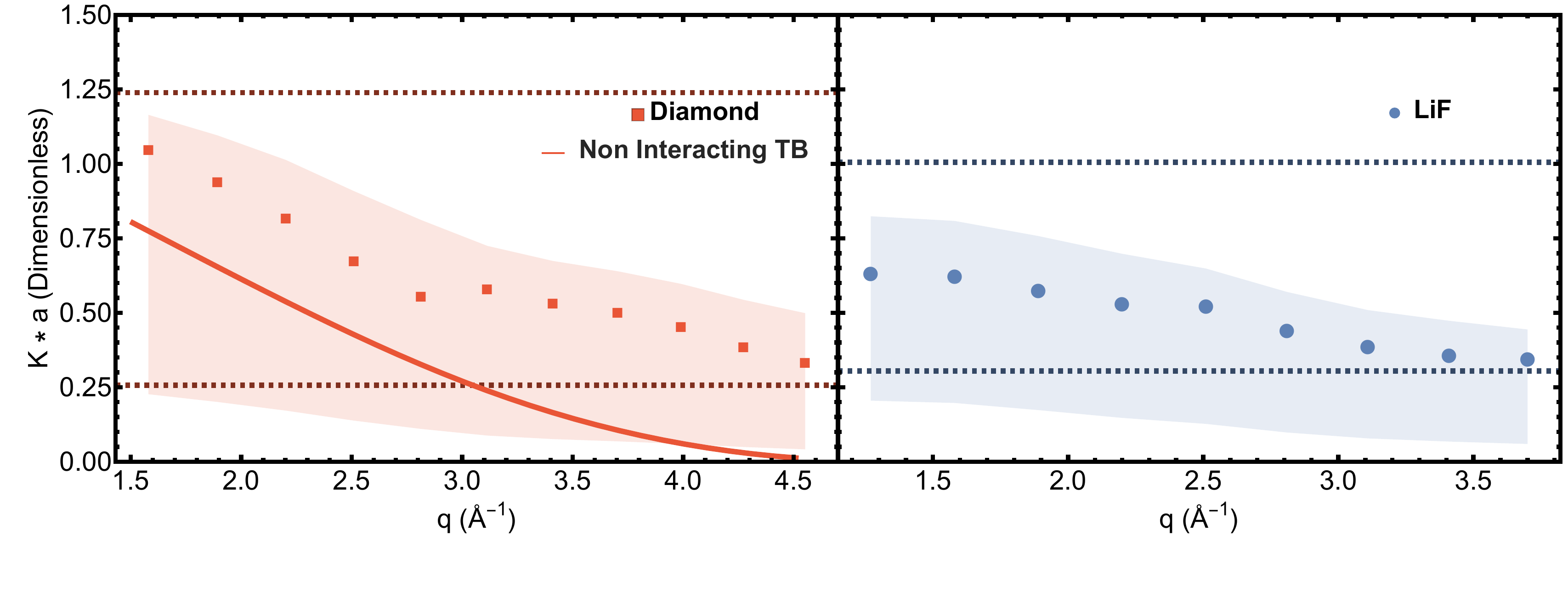}
        \caption{Values of $K(\mathbf{q})$ from IXS spectra. The shaded region is defined by the bounds Eq.~\eqref{eq: K Bound} with $\epsilon(\mathbf{q})$ from IXS data used and the $\mathbf{q} \rightarrow 0$ value of the gap used. Horizontal lines correspond to the $\mathbf{q \rightarrow 0}$ limit of Eq.~\eqref{eq: K Bound} using $\epsilon$ as reported in \cite{onishiQuantumWeightFundamental2025} and the same $n$ as used to scale the data with the f-sum rule.}
        \label{fig: K result}
    \end{figure}
\end{center}
\twocolumngrid\

We now combine the experimentally determined quantum Fisher information shown in Fig.~\ref{fig: QFI result} with the empirically extracted dielectric function to analyze the dimensionless quantum weight $aK(\mathbf{q})$ and its bounds for LiF and diamond. The results are summarized in Fig.~\ref{fig: K result}, which displays $aK(\mathbf{q})$ as a function of $q$ for diamond (blue) and LiF (orange). The shaded regions indicate the values permitted by the $\mathbf{q}$-dependent bounds of Eq.~\eqref{eq: K Bound} for each material. The dashed horizontal lines denote the zero-wavevector upper and lower bounds defined in Ref.~\cite{onishiQuantumWeightFundamental2025}.

Fig.~\ref{fig: K result} supports several conclusions. First, for all measured values of $q$, the dimensionless quantum weight for both materials lies within the bounds prescribed by Eq.~\eqref{eq: K Bound}. Extrapolating the trends in Fig.~\ref{fig: K result} to zero wavevector suggests that the quantum weight in LiF falls near the center of the $q=0$ bound of Ref. ~\cite{onishiQuantumWeightFundamental2025}, whereas in diamond it approaches the upper bound. This behavior is consistent with expectations that the valence electrons in diamond, a covalently bonded and obstructed atomic limit insulator~\cite{bradlyn2017topological,komissarov2024quantum,xu2021three}, are delocalized to the maximum degree possible given the size of its band gap. 

It is instructive to directly compare the dimensionless quantum weight $aK(q)$ in diamond and LiF. As emphasized above, $aK(q)$ quantifies the degree of electronic localization in the material. In the limit $q \rightarrow 0$, it provides a dimensionless measure of position fluctuations or, equivalently, of polarization density fluctuations in units of the primitive lattice constant. A value of $aK(q)<1$ indicates that electrons are largely confined within a single primitive unit cell, while $aK(q)>1$ implies that the ground-state electronic wave function extends across multiple unit cells. The quantity $aK(q)$ therefore serves as a quantitative measure of bond covalency.

Consistent with this interpretation, we find $aK(q\rightarrow 0)<1$ for the ionic insulator LiF and $aK(q\rightarrow 0)>1$ for diamond, a maximally covalent insulator. 
The value obtained for LiF is slightly smaller than in our previous study~\cite{balutQuantumEntanglementQuantum2025}, reflecting improved treatment of the elastic line in the present analysis.
These results demonstrate that inelastic x-ray scattering can be used to directly probe the nature of chemical bonding in insulating materials.

\section{Outlook}

One of the main conclusions of our study, based on the connection between the quantum weight and the Bures metric, is that there is a direct relationship between the covalency of bonding, as measured by $aK(q)$, and the degree of quantum entanglement in an insulator. 

To illustrate this point, in Appendix~\ref{sec:tight_binding_models} we compute the QFI and quantum weight for the one-dimensional SSH chain, an exactly solvable model that undergoes a transition from a trivial insulator to a covalently bonded (obstructed atomic) insulating phase, for which a lower bound on quantum entanglement can be computed exactly. As shown in Fig.~8, tuning the system through this transition reveals a clear correlation between the QFI and the entanglement lower bound, demonstrating a direct relationship between these quantities.

Our study therefore shows that electrons in diamond are not only more delocalized than in LiF, but also more strongly entangled, and that density perturbations in diamond imprint more information into its many-body quantum state. This enhanced response arises because a greater amount of quantum information is encoded in the covalent bonds of diamond, which are delocalized over multiple primitive unit cells. Together, these results reveal a deep connection between quantum information, electron localization, and chemical bonding, and demonstrate how this connection can be quantified experimentally using IXS. 

The increased information density in diamond is at least partly attributable to correlation effects. To illustrate this, in Appendix~\ref{sec:tight_binding_models} we compute the dimensionless quantum weight, $aK(\mathbf{q}) = 2\pi a S(\mathbf{q})/q^2$, using a tight-binding model of the $sp^3$ band structure of diamond. The result, shown as a solid line in Fig.~\ref{fig: K result}, systematically underestimates the measured value. This discrepancy arises because the tight-binding model neglects both electron–electron interactions and the finite spatial extent of the carbon atomic orbitals. Since these orbitals are small compared to the lattice constant, we expect correlation is the dominant effect leading to this discrepancy. This supports the intuition that interactions enhance entanglement and information density in a many-body wavefunction.
 
By connecting the quantum Fisher information and quantum weight to the Bures metric, our work shows that these quantities have a quantum-geometric interpretation. Furthermore at zero temperature they reduce to the \emph{longitudinal} component of the quantum metric. While it is the transverse component of the quantum metric that has received the most attention in the literature~\cite{yu2025quantum}, we have demonstrated that the longitudinal quantum metric is directly experimentally accessible and carries information about electron localization in solids.

A key advantage of the longitudinal Bures metric, Eq.~\eqref{eq:densitymetric}, or equivalently the longitudinal quantum weight $K(\mathbf{q})$, Eq.~\eqref{eq:Kqdef},
is that it is well-defined even in a metal. In metallic systems, we expect $K(\mathbf{q})$ to remain finite and to reflect the quantum Fisher information associated with screened density fluctuations. In this sense, $K(\mathbf{q})$ quantifies the distinguishability between the equilibrium metallic state and the state perturbed by a screened charge excitation, highlighting a direct connection between charge screening and quantum information.

In particular, we see that the dimensionless quantum weight, in encoding information about covalent bonds, is signaling the degree to which the electronic wavefunctions in an insulator are delocalized. Delocalization occurs through the sharing of the electronic wavefunction between basis Wannier orbitals localized on different atoms. As such, we expect the quantum Fisher information associated to density perturbations to encode the entanglement present in the ground state wavefunction between orbitals localized on different atoms. For noninteracting systems this can be made precise by considering the entanglement in the ground state between two subregions, where each subregion contains half the atoms (or more generally, half the orbitals) in each unit cell. Such translation-invariant partitions were considered for free fermion systems in Refs.~\cite{legnerRelatingEntanglementSpectrum2013,bradlyn2019disconnected}. By reinterpreting the results of Ref.~\cite{legnerRelatingEntanglementSpectrum2013} in a modern lens, we show in Appendix~\ref{sec:bounds_from_entanglement} that the eigenvalues of the reduced density matrix for the translation-invariant partition places a weak lower bound on the dimensionless quantum weight $aK(\mathbf{q})$, and hence on the Bures metric and quantum Fisher information. It is a pressing area for future theoretical work to explore whether a stronger bound on entanglement can be derived, and whether these results can be extended to interacting systems.

\begin{acknowledgments}
We gratefully acknowledge N. de Vries and D. Chaudhuri for helpful discussions, and T. Gog for experimental support. IXS experiments and data analysis were supported by the Center for Quantum Sensing
and Quantum Materials, an Energy Frontier Research Center funded by the
US Department of Energy (DOE), Office of Science, Basic Energy Sciences (BES), under award DE-SC0021238.
Theoretical work was supported by DOE BES grant no. DE-SC0026342. D.B. was supported in part by the Barry Goldwater Scholarship and the Astronaut Scholarship Foundation. 
P.A. gratefully acknowledges support from the EPiQS initiative of the Gordon and Betty Moore Foundation, grant GBMF9452. 
Use of the Advanced Photon Source was supported by the U.S. Department of Energy (DOE) Office of Science contract no. DE-AC02-06CH11357. 
\end{acknowledgments}

\appendix 
\section{Bounds from Entanglement}\label{sec:bounds_from_entanglement}
In this appendix, we explore the relationship between the quantum weight $K$ and entanglement for noninteracting electron systems. We begin with a general discrete translation-invariant tight-binding Hamiltonian of the form
\begin{equation}
    H = \sum_{i,j} c^\dagger_i h_{i,j} c_j,
    \label{eq: Tight Binding Hamiltonion}
\end{equation}
where $c^\dagger$,$c$ are single fermion creation and annihilation operators respectfully and $i,j$ index the degrees of freedom of the model (spin, orbital, and position).

Let us consider a partition of the degrees of freedom into two subsets in such a way  that each subset preserves the discrete translation symmetry of the model. We denote the projection operator onto the first subset as $\Pi_1$. The entanglement entropy associated with this partition is given by:
\begin{equation}
    S_1 = - \text{Tr}\left(\rho_1 \log \rho_1 \right)
    \label{eq: entanglement entropy/ von neuman entropy}
\end{equation}
where $\rho_1$ is the reduced density matrix associated with $\Pi_1$ and is defined in terms of the many-body ground state $|\boldsymbol{\Psi}\rangle$: 
\begin{equation}
    S_1 = \text{Tr}_2 |\boldsymbol{\Psi}\rangle \langle \boldsymbol{\Psi}|
\end{equation}
Here $\text{Tr}_2$ denotes a partial trace over the degrees of freedom not in $\Pi_1$. 

For noninteracting tight binding models, we can compute the reduced density matrix $\rho_1$ from a single-particle ``entanglement Hamiltonian" $H_1^e$ defined via~\cite{peschelCalculationReducedDensity2003}
\begin{equation}
    \rho_1 = \frac{e^{-H_1^e}}{\text{Tr}_1 e^{-H_1^e}} 
\end{equation}
Thus, the entanglement Hamiltonian can be written as~\cite{peschelCalculationReducedDensity2003}
\begin{equation}
    H^1_e = \sum_{i,j} c^\dagger_i h^e_{i,j} c_j.
    \label{eq: entanglement hamiltonian def}
\end{equation}
Crucially, because the partition $\Pi_1$ respects the translation symmetry of the Hamiltonian, the entanglement Hamiltonian $H_e^1$ is also translation invariant. We can thus index the eigenvalues of $h_e^1$ by crystal momentum $\mathbf{k}$. We denote by $\epsilon_a(\mathbf{k})$ the $a$-th eigenvalue of $h_e^1$

Ref.~\cite{legnerRelatingEntanglementSpectrum2013} showed that the diagonal components of the momentum-space quantum metric are lower bounded by the eigenvalues of $h^e_{i,j}$. Introducing $\lambda_a(\mathbf{k}) \equiv \tanh(\epsilon_a(\mathbf{k}/2)/2$ and $m \equiv \text{tr}\Pi_1$, we have
\begin{equation}
    g^{\mu\mu}(\mathbf{k}) \ge \frac{1}{n/2 - \sum \lambda_a(\mathbf{k})} \sum_{a,b}^m(\partial^\mu \lambda(\mathbf{k})_a)(\partial^\mu \lambda(\mathbf{k})_b),
    \label{eq: legner neupert relation between entanglement and geometry}
\end{equation}
where the partial derivatives are taken with respect to components of the crystal momentum. 

To connect this to the Bures metric, we recall that in the tight binding limit, the quantum Weight and momentum space quantum metric are related at zero temperature by~\cite{onishiQuantumWeightFundamental2025}
\begin{equation}\label{eq: K from gk}
    K^{\mu\nu} = \int_\text{BZ} \frac{d^d k}{(2 \pi)^d} g^{\mu\nu} (k).
\end{equation}
Combining Eqs.~\eqref{eq: legner neupert relation between entanglement and geometry} and \eqref{eq: K from gk} we arrive at an entanglement-based lower bound on the quantum weight, 
\begin{equation}
    K \ge  \int_\text{BZ} \frac{d^d k}{(2 \pi)^d} * \frac{1}{n/2 - \sum \lambda(k)_i} * \sum_{i,j}^m(\partial_\mu \lambda(k)_i)(\partial_\mu \lambda(k)_j)
    \label{eq: K bound from entanglement}
\end{equation}
Given Eqs.~\eqref{quantumweightdef} and \eqref{eq: QFI sum rule}, Eq.~\eqref{eq: K bound from entanglement} shows how the QFI can be interpreted as a measure of entanglement. Importantly, Eq.~\eqref{eq: K bound from entanglement} is valid for any partition $\Pi_1$ that preserves translation symmetry. 

For tight-binding models with a finite number of degrees of freedom per unit cell, we can also derive an upper bound on the QFI, and hence on the $\mathbf{q}$-dependent quantum weight.  Using Eqs.~\eqref{eq:densitymetric}  and \eqref{eq:zeroTssf} to write the QFI in terms of the static structure factor $S(\mathbf{q})$ of an insulator at zero temperature~\cite{onishiQuantumWeightFundamental2025}, we find
\begin{equation}
    f_Q = 4 S(\mathbf{q}) = 4 \int\frac{d^dk}{(2\pi)^d} \sum_{a <n}\sum_{b > n} |\langle u_{a,\mathbf{k}}|u_{b,\mathbf{k} + \mathbf{q}} \rangle|^2
    \label{eq: QFI in terms of projection operators}
\end{equation}
where $|u_a,\mathbf{k}\rangle$ is the normalized Bloch state of the $a$-th band at momentum $\mathbf{k}$ and $n$ is the number of occupied bands. Using Cauchy-Schwarz we have that
\begin{equation}\label{eq:qfiupperbound}
   4 n(N-n) \ge f_Q
\end{equation}
where $N$ is the total number of bands. We note that while the lower bound Eq.~\eqref{eq: K bound from entanglement} is valid only in the limit $\mathbf{q}\rightarrow 0$, the upper bound Eq.~\eqref{eq:qfiupperbound} is valid for any $\mathbf{q}$.

\section{Tight Binding Models}\label{sec:tight_binding_models}
Here we explore the quantum weight (and hence QFI and Bures metric), as well as Eq.~\eqref{eq: K bound from entanglement} in the context of common tight binding models. Though Eq.~\eqref{eq: K bound from entanglement} provides a relationship between entanglement and geometry, Ref.~\cite{legnerRelatingEntanglementSpectrum2013} showed that one can write the left hand side in terms of projection operators onto occupied bands $\Pi(\mathbf{k}) \equiv \sum_{i} \ |u_{i,\mathbf{k}}\rangle\langle u_{i,\mathbf{k}}|$ and the translationally invariant partition $\Pi_1$: 

\begin{widetext}
    \begin{equation}
        \frac{1}{n/2 - \sum \lambda_a(\mathbf{k})} * \sum_{a,b}^m(\partial^\mu \lambda_a(\mathbf{k}))(\partial^\mu \lambda_b(\mathbf{k})) = \left(\frac{m + n}{2} - \text{Tr}(\Pi_1 \Pi(\mathbf{k})\Pi_1) \right)  * \left(\partial^\mu \log \left( \sqrt{\frac{m + n}{2} - \text{Tr}(\Pi_1 \Pi(\mathbf{k})\Pi_1)} \right) \right)^2
        \label{eq: lower bound in terms of partition}
    \end{equation}    
\end{widetext}

Going further, we can use Eq.~\eqref{Kdef} to write the QFI and quantum weight in the small $\mathbf{q}$ limit in terms of the static structure factor at zero temperature,
\begin{equation}
    \frac{2}{\pi} q_\mu q_\nu K^{\mu\nu} =f_Q(\mathbf{q}\rightarrow 0) =  4 S(\mathbf{q}\rightarrow 0)
    \label{eq: Relate QFI to K and Sq}
\end{equation}
For a noninteracting insulator, we can use the occupied Bloch wavefunctions to write $S(\mathbf{q})$ as ~\cite{onishiQuantumWeightFundamental2025}
\begin{equation}
    S(\mathbf{q}) = \int_{\text{BZ}}\frac{d^d\mathbf{k}}{(2\pi)^d}\text{Tr}[P(\mathbf{k})(P(\mathbf{k}) - P(\mathbf{k} + \mathbf{q})]
    \label{eq: Sq Def}
\end{equation}
We will apply Eqs.~\eqref{eq: Relate QFI to K and Sq} and \eqref{eq: Sq Def} to study the QFI and quantum weight for two illustrative tight-binding models. First, in Sec.~\ref{sec:ssh} we will look at how the QFI varies across the topological phase transition in the Su-Schrieffer-Heeger (SSH) model; we will interpret these results in terms of a change in covalent bond character. Next, in Sec.~\ref{sec:diamond} we will compute the dimensionless quantum weight for an empirical 8-band tight-binding model of diamond, and compare with our experimental results.

\subsection{SSH Model}\label{sec:ssh}
The SSH Bloch Hamiltonian takes the form
\begin{subequations}
    \begin{align}
       & H = \mathbf{d}\cdot\mathbf{\sigma} \\ 
        & \mathbf{d} = \left(0,t \sin(k a), \Delta + t \cos(k a)\right),
    \end{align}
\end{subequations}

\onecolumngrid\
\begin{center}\
\begin{figure}
    \centering
    \includegraphics[width=\linewidth]{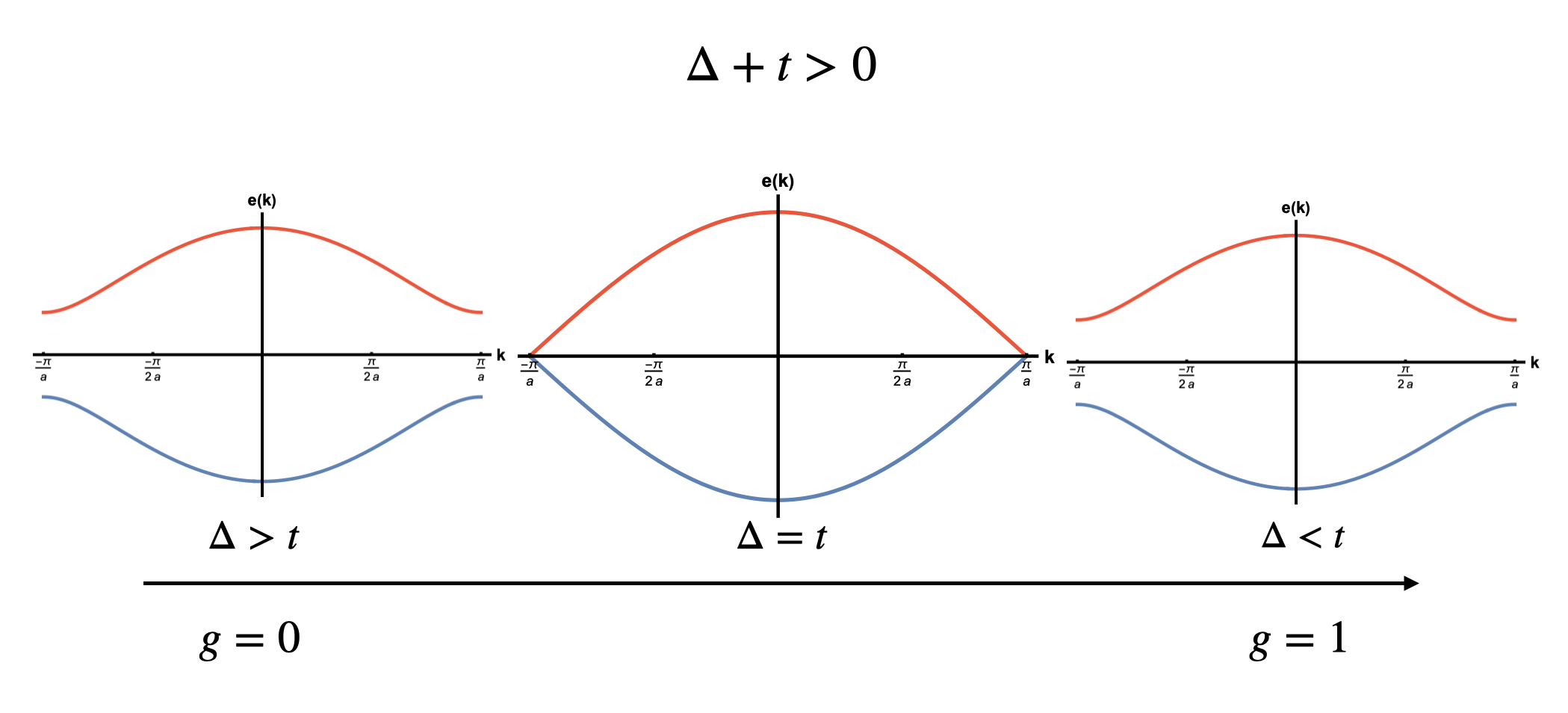}
    \caption{The phases of the SSH model for $\Delta + t >0$. $\pi g$ is the Berry phase of the occupied bands, which is quantized due to inversion symmetry. $g=0$ corresponds to the trivial atomic insulator, while $g=1$ corresponds to the nontrivial obstructed atomic phase.}
    \label{fig: SSH phase diagram}
\end{figure}
\end{center}
\twocolumngrid\

\begin{figure}
    \centering
    \includegraphics[width=\linewidth]{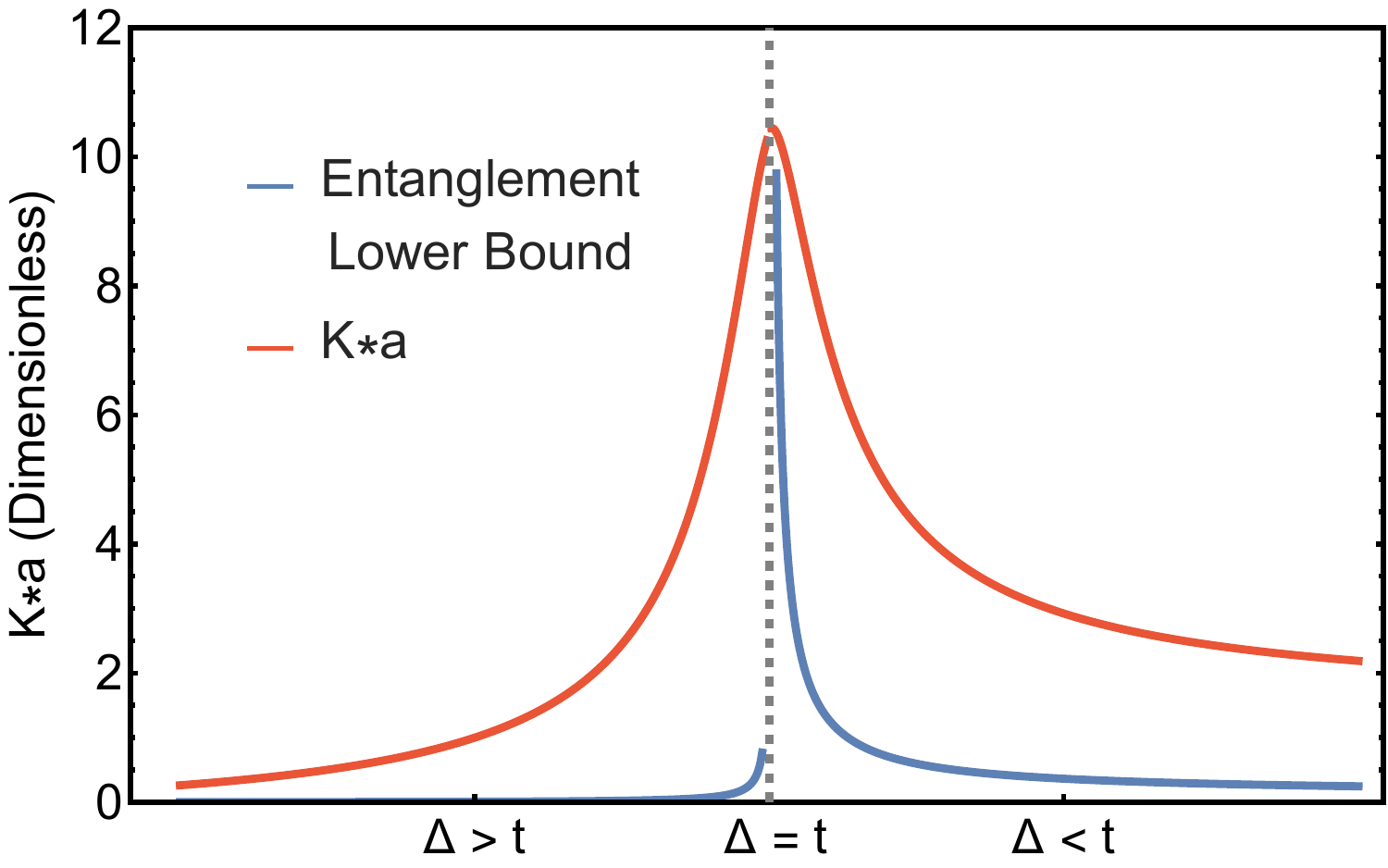}
    \caption{Comparison between $aK(q)$ from the SSH model and the entanglement lower bound Eq.~\eqref{eq: K bound from entanglement}. Here we partition the "s-like" from the "p-like" orbital. The comparison was made for $q = .1$ with lattice spacing set to one.}
    \label{fig: SSH QFI and Bound Result}
\end{figure}

To investigate this connection, we consider the partition \begin{equation}
    \Pi_1  = 
\begin{pmatrix}
1 & 0 \\
0 & 0
\end{pmatrix}
\end{equation}
which partitions the $s$ orbital from the $p$ orbital. Then here, $m = \text{Tr}\Pi_1 = 1$ and $n = 1$ the number of occupied bands. Then, using Eqs.~\eqref{eq: K bound from entanglement},~\eqref{eq: Relate QFI to K and Sq},~\eqref{eq: lower bound in terms of partition} we can numerically compute a lower bound on the QFI originating from entanglement between the $s$ and $p$ orbital degrees of freedom. Finally, we use Eqs.~\eqref{eq: Relate QFI to K and Sq},~\eqref{eq: Sq Def} to compute the QFI of the SSH model. We show the results of the computation in Fig.~\ref{fig: SSH QFI and Bound Result}. Importantly, for $\Delta < t$ the QFI asymptotically approaches a nonzero value as $\Delta\rightarrow 0$ which can be attributed to the delocalization of Wannier functions. It reflects the quantum (Fisher) information shared between adjacent unit cells in the model.The SSH model thus shows that the QFI indeed captures electronic delocalization in insulators.

\subsection{Diamond Tight Binding Model}\label{sec:diamond}
We now move to an 8 orbital tight binding model for diamond. Values of hopping terms and the $8 \times 8$ Hamiltonian are found in Ref.~\cite{chadi_tight-binding_1975}. The basis functions of the model are the eight $sp^3$ hybrid orbitals on the two carbon atoms in each diamond unit cell.  Once again, we use Eq.~\eqref{eq: Sq Def} to numerically compute the structure factor. In Fig.~\ref{fig: K result} we show in red the dimensionless quantum weight $K(\mathbf{q})a = 2\pi a S(\mathbf{q})/q^2$ as a function of $\mathbf{q}$, for $\mathbf{q} = q/\sqrt{2} (1,-1,0)$. We find that the tight-binding model qualitatively captures the momentum dependence of $K$ seen in the experimental data. As expected, the tight-binding model under estimates the value of $K$ measured experimentally. This can be attributed to several sources of delocalization that are present in the real material but not in the model, such as the finite spread of atomic orbitals and correlation effects. The accuracy of the model prediction increases for small $q$, indicating that model provides an accurate estimate of interband contributions to the quantum weight. 

We can also compute the lower bound Eq.~\eqref{eq: K bound from entanglement} originating from entanglement just as in the SSH model. We consider a partition along the direction $\mathbf{q} = q/\sqrt{2} (1,-1,0)$. We partition the s orbitals from the p orbitals with the projection operator in the basis given by Ref.~\cite{chadi_tight-binding_1975}:
\begin{equation}
   \Pi_{sp} =  \mathrm{diag}(1,1,0,0,0,0,0,0)
\end{equation}
 Doing so gives the lower bound $K*a \ge 4.94 \times10^{-4}$. 

\bibliography{References}

@online{PrecisionMeasurementLattice,
  title = {Precision Measurement of Lattice Parameters in {{LiF}} Monocrystals - {{Dressler}} - 1987 - {{Crystal Research}} and {{Technology}} - {{Wiley Online Library}}},
  url = {https://onlinelibrary.wiley.com/doi/abs/10.1002/crat.2170221116?casa_token=hwtHxSnoNAkAAAAA%3AvfKue-IvQ7H0S--PxJA3iNXs9BtHysvrrqAg6VmGlH62ciipvxuHOcmXxPY5acEdl1BXG9tFr71b3BNg},
  urldate = {2025-07-18}
}

@article{balutQuantumEntanglementQuantum2025,
  title = {Quantum Entanglement and Quantum Geometry Measured with Inelastic X-Ray Scattering},
  author = {Bałut, David and Bradlyn, Barry and Abbamonte, Peter},
  date = {2025-03-31},
  journal = {Physical Review B},
  shortjournal = {Phys. Rev. B},
  volume = {111},
  number = {12},
  pages = {125161},
  publisher = {American Physical Society},
  doi = {10.1103/PhysRevB.111.125161},
  url = {https://link.aps.org/doi/10.1103/PhysRevB.111.125161},
  urldate = {2025-04-15}
}

@article{balut2025quantum,
  title={Quantum fisher information reveals UV-IR mixing in the strange metal},
  author={Ba{\l}ut, David and Guo, Xuefei and de Vries, Niels and Chaudhuri, Dipanjan and Bradlyn, Barry and Abbamonte, Peter and Phillips, Philip W},
  journal={Physica C: Superconductivity and its Applications},
  volume={635},
  pages={1354750},
  year={2025},
  publisher={Elsevier}
}

@online{guanExploringManyBodyQuantum2025,
  title = {Exploring {{Many-Body Quantum Geometry Beyond}} the {{Quantum Metric}} with {{Correlation Functions}}: {{A Time-Dependent Perspective}}},
  shorttitle = {Exploring {{Many-Body Quantum Geometry Beyond}} the {{Quantum Metric}} with {{Correlation Functions}}},
  author = {Guan, Yuntao and Bradlyn, Barry},
  date = {2025-07-30},
  eprint = {2507.23028},
  eprinttype = {arXiv},
  eprintclass = {cond-mat},
  doi = {10.48550/arXiv.2507.23028},
  url = {http://arxiv.org/abs/2507.23028},
  urldate = {2025-08-06},
  pubstate = {prepublished}
}

@article{haukeMeasuringMultipartiteEntanglement2016,
  title = {Measuring Multipartite Entanglement through Dynamic Susceptibilities},
  author = {Hauke, Philipp and Heyl, Markus and Tagliacozzo, Luca and Zoller, Peter},
  date = {2016-08},
  journal = {Nature Physics},
  shortjournal = {Nature Phys},
  volume = {12},
  number = {8},
  pages = {778--782},
  publisher = {Nature Publishing Group},
  issn = {1745-2481},
  doi = {10.1038/nphys3700},
  url = {https://www.nature.com/articles/nphys3700},
  urldate = {2024-06-18},
  langid = {english}
}

@article{onishi2024fundamental,
  title={Fundamental bound on topological gap},
  author={Onishi, Yugo and Fu, Liang},
  journal={Physical Review X},
  volume={14},
  number={1},
  pages={011052},
  year={2024},
  publisher={APS}
}

@article{cordero2008covalent,
  title={Covalent radii revisited},
  author={Cordero, Beatriz and G{\'o}mez, Ver{\'o}nica and Platero-Prats, Ana E and Rev{\'e}s, Marc and Echeverr{\'\i}a, Jorge and Cremades, Eduard and Barrag{\'a}n, Flavia and Alvarez, Santiago},
  journal={Dalton Transactions},
  number={21},
  pages={2832--2838},
  year={2008},
  publisher={Royal Society of Chemistry}
}

@article{bradlyn2017topological,
  title={Topological quantum chemistry},
  author={Bradlyn, Barry and Elcoro, Luis and Cano, Jennifer and Vergniory, Maia G and Wang, Zhijun and Felser, Claudia and Aroyo, Mois I and Bernevig, B Andrei},
  journal={Nature},
  volume={547},
  number={7663},
  pages={298--305},
  year={2017},
  publisher={Nature Publishing Group UK London}
}

@article{onishiQuantumWeightFundamental2025,
  title = {Quantum Weight: {{A}} Fundamental Property of Quantum Many-Body Systems},
  shorttitle = {Quantum Weight},
  author = {Onishi, Yugo and Fu, Liang},
  date = {2025-05-19},
  journal = {Physical Review Research},
  shortjournal = {Phys. Rev. Res.},
  volume = {7},
  number = {2},
  pages = {023158},
  publisher = {American Physical Society},
  doi = {10.1103/PhysRevResearch.7.023158},
  url = {https://link.aps.org/doi/10.1103/PhysRevResearch.7.023158},
  urldate = {2025-06-04}
}

@article{komissarov2024quantum,
  title={The quantum geometric origin of capacitance in insulators},
  author={Komissarov, Ilia and Holder, Tobias and Queiroz, Raquel},
  journal={Nature communications},
  volume={15},
  number={1},
  pages={4621},
  year={2024},
  publisher={Nature Publishing Group UK London}
}

@article{xu2021three,
  title={Three-dimensional real space invariants, obstructed atomic insulators and a new principle for active catalytic sites},
  author={Xu, Yuanfeng and Elcoro, Luis and Li, Guowei and Song, Zhi-Da and Regnault, Nicolas and Yang, Qun and Sun, Yan and Parkin, Stuart and Felser, Claudia and Bernevig, B Andrei},
  journal={arXiv preprint arXiv:2111.02433},
  year={2021}
}

@article{bradlyn2019disconnected,
  title={Disconnected elementary band representations, fragile topology, and Wilson loops as topological indices: An example on the triangular lattice},
  author={Bradlyn, Barry and Wang, Zhijun and Cano, Jennifer and Bernevig, B Andrei},
  journal={Physical Review B},
  volume={99},
  number={4},
  pages={045140},
  year={2019},
  publisher={APS}
}

@article{legnerRelatingEntanglementSpectrum2013,
  title = {Relating the Entanglement Spectrum of Noninteracting Band Insulators to Their Quantum Geometry and Topology},
  author = {Legner, Markus and Neupert, Titus},
  date = {2013-09-09},
  journal = {Physical Review B},
  shortjournal = {Phys. Rev. B},
  volume = {88},
  number = {11},
  pages = {115114},
  publisher = {American Physical Society},
  doi = {10.1103/PhysRevB.88.115114},
  url = {https://link.aps.org/doi/10.1103/PhysRevB.88.115114},
  urldate = {2025-08-13}
}

@article{peschelCalculationReducedDensity2003,
  title = {Calculation of Reduced Density Matrices from Correlation Functions},
  author = {Peschel, Ingo},
  date = {2003-03},
  journal = {Journal of Physics A: Mathematical and General},
  shortjournal = {J. Phys. A: Math. Gen.},
  volume = {36},
  number = {14},
  pages = {L205},
  issn = {0305-4470},
  doi = {10.1088/0305-4470/36/14/101},
  url = {https://dx.doi.org/10.1088/0305-4470/36/14/101},
  urldate = {2025-08-13},
  langid = {english}
}

@article{straumanis_precision_1951,
    title = {Precision {Determination} of {Lattice} {Parameter}, {Coefficient} of {Thermal} {Expansion} and {Atomic} {Weight} of {Carbon} in {Diamond1}},
    volume = {73},
    issn = {0002-7863},
    url = {https://doi.org/10.1021/ja01156a043},
    doi = {10.1021/ja01156a043},
    number = {12},
    urldate = {2025-08-06},
    journal = {Journal of the American Chemical Society},
    author = {Straumanis, M. E. and Aka, E. Z.},
    month = dec,
    year = {1951},
    note = {Publisher: American Chemical Society},
    pages = {5643--5646},
}

@article{chadi_tight-binding_1975,
    title = {Tight-binding calculations of the valence bands of diamond and zincblende crystals},
    volume = {68},
    copyright = {Copyright © 1975 WILEY-VCH Verlag GmbH \& Co. KGaA},
    issn = {1521-3951},
    url = {https://onlinelibrary.wiley.com/doi/abs/10.1002/pssb.2220680140},
    doi = {10.1002/pssb.2220680140},
    number = {1},
    urldate = {2025-08-06},
    journal = {physica status solidi (b)},
    author = {Chadi, D. J. and Cohen, M. L.},
    year = {1975},
    note = {\_eprint: https://onlinelibrary.wiley.com/doi/pdf/10.1002/pssb.2220680140},
    pages = {405--419},
}

@article{niu1985quantized,
  title = {Quantized {{Hall}} Conductance as a Topological Invariant},
  author = {Niu, Qian and Thouless, Ds J and Wu, Yong-Shi},
  year = {1985},
  journal = {Physical Review B},
  volume = {31},
  number = {6},
  pages = {3372},
  publisher = {APS}
}

@article{provost1980riemannian,
  title = {Riemannian Structure on Manifolds of Quantum States},
  author = {Provost, {\relax JP} and Vallee, G},
  year = {1980},
  journal = {Communications in Mathematical Physics},
  volume = {76},
  pages = {289--301},
  publisher = {Springer}
}

@article{souza2000polarization,
  title = {Polarization and Localization in Insulators: {{Generating}} Function Approach},
  author = {Souza, Ivo and Wilkens, Tim and Martin, Richard M},
  year = {2000},
  journal = {Physical Review B},
  volume = {62},
  number = {3},
  pages = {1666},
  publisher = {APS}
}

@article{ozawa2021relations,
  title = {Relations between Topology and the Quantum Metric for {{Chern}} Insulators},
  author = {Ozawa, Tomoki and Mera, Bruno},
  year = {2021},
  journal = {Physical Review B},
  volume = {104},
  number = {4},
  pages = {045103},
  publisher = {APS}
}

@book{schuelkeElectronDynamicsInelastic2007,
  title = {Electron {{Dynamics}} by {{Inelastic X-Ray Scattering}}},
  author = {Schuelke, Winfried},
  year = {2007},
  series = {Oxford {{Series}} on {{Synchrotron Radiation}}},
  publisher = {Oxford University Press},
  location = {Oxford, New York},
  isbn = {978-0-19-851017-8},
  pagetotal = {606}
}

@article{yu2025quantum,
  title = {Quantum Geometry in Quantum Materials},
  author = {Yu, Jiabin and Bernevig, B Andrei and Queiroz, Raquel and Rossi, Enrico and T{\"o}rm{\"a}, P{\"a}ivi and Yang, Bohm-Jung},
  year = 2025,
  journal = {npj Quantum Materials},
  volume = {10},
  number = {1},
  pages = {101},
  publisher = {Nature Publishing Group UK London}
}

@article{marzari1997maximally,
  title = {Maximally Localized Generalized {{Wannier}} Functions for Composite Energy Bands},
  author = {Marzari, Nicola and Vanderbilt, David},
  year = 1997,
  journal = {Physical review B},
  volume = {56},
  number = {20},
  pages = {12847},
  publisher = {APS}
}

@article{parameswaran2013fractional,
  title = {Fractional Quantum {{Hall}} Physics in Topological Flat Bands},
  author = {Parameswaran, Siddharth A and Roy, Rahul and Sondhi, Shivaji L},
  year = 2013,
  journal = {Comptes Rendus Physique},
  volume = {14},
  number = {9-10},
  pages = {816--839},
  publisher = {Elsevier}
}

@article{torma2023essay,
  title = {Essay: {{Where}} Can Quantum Geometry Lead Us?},
  author = {T{\"o}rm{\"a}, P{\"a}ivi},
  year = 2023,
  journal = {Physical Review Letters},
  volume = {131},
  number = {24},
  pages = {240001},
  publisher = {APS}
}

@article{torma2022superconductivity,
  title = {Superconductivity, Superfluidity and Quantum Geometry in Twisted Multilayer Systems},
  author = {T{\"o}rm{\"a}, P{\"a}ivi and Peotta, Sebastiano and Bernevig, Bogdan A},
  year = 2022,
  journal = {Nature Reviews Physics},
  volume = {4},
  number = {8},
  pages = {528--542},
  publisher = {Nature Publishing Group UK London}
}

@article{robinson1997reinterpretation,
  title={Reinterpretation of the lengths of bonds to fluorine in terms of an almost ionic model},
  author={Robinson, Edward A and Johnson, Samuel A and Tang, Ting-Hua and Gillespie, Ronald J},
  journal={Inorganic chemistry},
  volume={36},
  number={14},
  pages={3022--3030},
  year={1997},
  publisher={ACS Publications}
}

@article{ji2025density,
  title = {Density Matrix Geometry and Sum Rules},
  author = {Ji, Guangyue and Palomino, David E and Goldman, Nathan and Ozawa, Tomoki and Riseborough, Peter and Wang, Jie and Mera, Bruno},
  year = 2025,
  journal = {arXiv preprint arXiv:2507.14028},
  eprint = {2507.14028},
  archiveprefix = {arXiv}
}

@article{verma2025quantum,
  title = {Quantum Geometry: {{Revisiting}} Electronic Scales in Quantum Matter},
  author = {Verma, Nishchhal and Moll, Philip JW and Holder, Tobias and Queiroz, Raquel},
  year = 2025,
  journal = {arXiv preprint arXiv:2504.07173},
  eprint = {2504.07173},
  archiveprefix = {arXiv}
}

@book{Bengtsson_Zyczkowski_2006,
  title = {Geometry of Quantum States: {{An}} Introduction to Quantum Entanglement},
  author = {Bengtsson, Ingemar and Zyczkowski, Karol},
  year = 2006,
  publisher = {Cambridge University Press},
  address = {Cambridge}
}

@article{uhlmann1976transition,
  title={The ``transition probability'' in the state space of a $\ast$-algebra},
  author={Uhlmann, Armin},
  journal={Reports on Mathematical Physics},
  volume={9},
  number={2},
  pages={273--279},
  year={1976},
  publisher={Elsevier}
}

@article{braunstein1994statistical,
  title = {Statistical Distance and the Geometry of Quantum States},
  author = {Braunstein, Samuel L. and Caves, Carlton M.},
  year = 1994,
  month = may,
  journal = {Physical Review Letters},
  volume = {72},
  number = {22},
  pages = {3439--3443},
  publisher = {American Physical Society},
  doi = {10.1103/PhysRevLett.72.3439},
  url = {https://link.aps.org/doi/10.1103/PhysRevLett.72.3439}
}

@article{hetenyi2023fluctuations,
  title = {Fluctuations, Uncertainty Relations, and the Geometry of Quantum State Manifolds},
  author = {Het{\'e}nyi, Bal{\'a}zs and L{\'e}vay, P{\'e}ter},
  year = 2023,
  journal = {Physical Review A},
  volume = {108},
  number = {3},
  pages = {032218},
  publisher = {APS}
}

@article{kadanoff1963hydrodynamic,
  title = {Hydrodynamic Equations and Correlation Functions},
  author = {Kadanoff, Leo P and Martin, Paul C},
  year = 1963,
  journal = {Annals of Physics},
  volume = {24},
  pages = {419--469},
  publisher = {Elsevier}
}

@article{resta2006polarization,
  title = {Polarization Fluctuations in Insulators and Metals: New and Old Theories Merge},
  author = {Resta, Raffaele},
  year = 2006,
  journal = {Physical review letters},
  volume = {96},
  number = {13},
  pages = {137601},
  publisher = {APS}
}

@article{AbbamonteLiF2008,
    author = {Peter Abbamonte  and Tim Graber  and James P. Reed  and Serban Smadici  and Chen-Lin Yeh  and Abhay Shukla  and Jean-Pascal Rueff  and Wei Ku },
    title = {Dynamical reconstruction of the exciton in LiF with inelastic x-ray scattering},
    journal = {Proceedings of the National Academy of Sciences},
    volume = {105},
    number = {34},
    pages = {12159-12163},
    year = {2008},
    doi = {10.1073/pnas.0801623105},
    URL = {https://www.pnas.org/doi/abs/10.1073/pnas.0801623105},
    eprint = {https://www.pnas.org/doi/pdf/10.1073/pnas.0801623105}
}

@article{Lee-LiF-2013,
  title = {First-Principles Method of Propagation of Tightly Bound Excitons: Verifying the Exciton Band Structure of LiF with Inelastic x-Ray Scattering},
  author = {Lee, Chi-Cheng and Chen, Xiaoqian M. and Gan, Yu and Yeh, Chen-Lin and Hsueh, H. C. and Abbamonte, Peter and Ku, Wei},
  journal = {Phys. Rev. Lett.},
  volume = {111},
  issue = {15},
  pages = {157401},
  numpages = {5},
  year = {2013},
  month = {Oct},
  publisher = {American Physical Society},
  doi = {10.1103/PhysRevLett.111.157401},
  url = {https://link.aps.org/doi/10.1103/PhysRevLett.111.157401}
}

@article{Caliebe2000,
  title = {Dynamic Structure Factor of Diamond and LiF Measured Using Inelastic X-Ray Scattering},
  author = {Caliebe, W. A. and Soininen, J. A. and Shirley, Eric L. and Kao, C.-C. and H\"am\"al\"ainen, K.},
  journal = {Phys. Rev. Lett.},
  volume = {84},
  issue = {17},
  pages = {3907--3910},
  numpages = {0},
  year = {2000},
  month = {Apr},
  publisher = {American Physical Society},
  doi = {10.1103/PhysRevLett.84.3907},
  url = {https://link.aps.org/doi/10.1103/PhysRevLett.84.3907}
}

@book{Martin1968,
  title = {Measurements and Correlation Functions},
  author = {Paul C. Martin},
  year = {1968},
  publisher = {Gordon and Breach Science Publishers},
  location = {New York, NY},
}

@article{Taft1965,
  title = {Optical Properties of Graphite},
  author = {Taft, E. A. and Philipp, H. R.},
  journal = {Phys. Rev.},
  volume = {138},
  issue = {1A},
  pages = {A197--A202},
  numpages = {0},
  year = {1965},
  month = {Apr},
  publisher = {American Physical Society},
  doi = {10.1103/PhysRev.138.A197},
  url = {https://link.aps.org/doi/10.1103/PhysRev.138.A197}
}

@article{Ehrenreich1963,
  title = {Optical Properties of Aluminum},
  author = {Ehrenreich, H. and Philipp, H. R. and Segall, B.},
  journal = {Phys. Rev.},
  volume = {132},
  issue = {5},
  pages = {1918--1928},
  numpages = {0},
  year = {1963},
  month = {Dec},
  publisher = {American Physical Society},
  doi = {10.1103/PhysRev.132.1918},
  url = {https://link.aps.org/doi/10.1103/PhysRev.132.1918}
}

@article{Shiles1980,
  title = {Self-consistency and sum-rule tests in the Kramers-Kronig analysis of optical data: Applications to aluminum},
  author = {Shiles, E. and Sasaki, Taizo and Inokuti, Mitio and Smith, D. Y.},
  journal = {Phys. Rev. B},
  volume = {22},
  issue = {4},
  pages = {1612--1628},
  numpages = {0},
  year = {1980},
  month = {Aug},
  publisher = {American Physical Society},
  doi = {10.1103/PhysRevB.22.1612},
  url = {https://link.aps.org/doi/10.1103/PhysRevB.22.1612}
}

@article{wang2025local,
  title = {Local and Non-Local Entanglement Witnesses of Fermi Liquid},
  author = {Wang, Yiming and Fang, Yuan and Xie, Fang and Si, Qimiao},
  year = {2025},
  journal = {arXiv preprint arXiv:2502.13958},
  eprint = {2502.13958},
  archiveprefix = {arXiv}
}

@article{mazza2024quantum,
  title={Quantum Fisher information in a strange metal},
  author={Mazza, Federico and Biswas, Sounak and Yan, Xinlin and Prokofiev, Andrey and Steffens, Paul and Si, Qimiao and Assaad, Fakher F and Paschen, Silke},
  journal={arXiv preprint arXiv:2403.12779},
  year={2024}
}

@article{fang2025amplified,
  title={Amplified multipartite entanglement witnessed in a quantum critical metal},
  author={Fang, Yuan and Mahankali, Mounica and Wang, Yiming and Chen, Lei and Hu, Haoyu and Paschen, Silke and Si, Qimiao},
  journal={Nature Communications},
  volume={16},
  number={1},
  pages={2498},
  year={2025},
  publisher={Nature Publishing Group UK London}
}

\end{document}